# A feasibility Study of Time of Flight Computed Tomography for Breast Imaging

Ignacio O. Romero, and Changqing Li*
Department of Bioengineering, University of California, Merced, Merced, CA, USA.

## Abstract

Cone beam computed tomography (CBCT) for breast imaging has potential to replace conventional mammograms. However, concerns over dose and image quality prevent CBBCT systems from the clinical trial phase to next stage. The time of flight (TOF) method was recently shown to reduce the x-ray scattering effects by 95% and improve the image CNR by 110% for large volume objects. The advancements in x-ray sources like in compact Free Electron Lasers (FEL) and advancements in detector technology show potential for the TOF method to be feasible in CBCT when imaging large objects. In this study, we investigate the efficacy of this TOF CBCT in improving the breast cancer imaging. The GATE software was used to simulate the cone beam CT imaging of an 8 cm diameter cylindrical water phantom using a modeled 20 keV quasi-energetic FEL source and various detector temporal resolutions ranging from 1 to 1000 ps. An inhomogeneous breast phantom of similar size was also imaged using the same system setup. Results show that a detector temporal resolution of 10 ps improved the image contrast-to-noise ratio (CNR) by 57% and reduced the scatter-to-primary ratio (SPR) by 8.63 for a small cylindrical phantom. For the breast phantom, the image CNR was enhanced by 12% and the SPR was reduced by 1.35 at 5 ps temporal resolution.

*cli32@ucmerced.edu; http://biomedimaging.ucmerced.edu

## 1. Introduction

In the United States of America, breast cancer ranks second in all cancers in terms of cancer mortality in the female population [1]. Mammography has remained as the standard for breast cancer imaging and has reduced the mortality rates of breast cancer [2]. However, mammography has relatively lower sensitivity and specificity in dense breasts due to tissue overlap that contributes to the masking of hidden cancers and the detection of false lesions [3].

Cone beam computed tomography (CBCT) has been established as a potential candidate to improve breast imaging. CBCT allows for three-dimensional imaging from chest wall to nipple of the breast without compression, and with radiation dose comparable to conventional mammography [4]. However, CBCT is susceptible to greater scatter noise due to the size of irradiated volume [5,6]. The scatter counts create inaccuracies in the reconstructed images that lead to a decrease of sensitivity in the detection of malignant tissues. Nevertheless, CBCT offers a promising modality for the evaluation of breast tissues due to its improved patient comfort and full field-of-view in evaluating lesion margins [7].

The rise of novel monochromatic x-ray sources shows promise in x-ray imaging. Achterhold et al compared monochromatic vs polychromatic x-ray tomographic imaging on a phantom sample where the findings confirmed that the monochromatic x-ray source can yield much higher CT image quality [8]. Unlike conventional polychromatic x-ray sources, free electron lasers (FELs) produce quasi-monoenergetic x-rays through a process known as inverse Compton scattering where an electromagnetic wave is amplified through interactions with a bunch of relativistic electrons [9]. The first FEL is based on a 2-mile-long Linac located at the SLAC National Accelerator Laboratory, Stanford, CA. The generated x-rays have unique features such as quasi-monochromatic characteristics, tunable energy, super short pulse, and coherence. These features make this new type of x-ray attractive for breast imaging. The x-ray energy of the compact FEL can be tuned to select the optimal monochromatic x-rays for breasts with different densities. The development of High-Gain Harmonic Generation (HGHG) FEL makes it possible to build a compact FEL which means it is possible to introduce them in future breast cancer imaging [10-13]. The super short x-ray pulse from FEL makes it possible to perform time of flight CT (TOF-CT) imaging, in which most scattered x-ray photons can be removed by analyzing the x-ray flight time so that the contrast-to-noise ratio (CNR) can be improved. The time of flight (TOF) CBCT has been previously introduced and demonstrated using a 24 cm diameter cylindrical water phantom with bone targets where up to 95% of the scatter counts were removed [14].

Detectors with a high time resolution are of high interest to improve image quality. An example is the time of fight detectors in PET [15,16]. In TOF PET, time-to-digital converters (TDC) are used to extract the difference between the arrival times of the start and stop signals to reconstruct the annihilation event. TOF PET studies have demonstrated that a detector time resolution of 500 picoseconds (ps) has a signal-to-noise ratio (SNR) improvement of about 2.3 times and sensitivity increase of about 5.3 times [16-19]. Gola et al reported a Fondazione Bruno Kessler (FBK) SiPM with a Lutetium Oxyorthosilicate crystal doped with Ce and Ca had achieved a time resolution of about 75 ps [20]. These recent developments have advanced TOF-PET imaging but there is no TOF detection system dedicated for CT. Recently, Cheng at al reported the time resolution of field programmable gate array (FPGA) TDC to be about 10 ps which makes it feasible to develop a TOF detector with a time resolution of 50 ps if only comparing the rising edge of the measurement signal [21].

In this paper, a TOF CBCT study is presented to investigate the efficacy of TOF CBCT in improving breast imaging in terms of dose reduction and contrast-to-noise ratio (CNR). The proposed TOF CBCT was simulated with GEANT4 Application of Tomographic Emission (GATE) software. GATE versions 8.0+ include newly featured time of flight measurements of each detected photon [22]. By extracting the time difference between detection and emission of the x-ray photon, the scatter noise counts can be separated from primary counts and thus improve the image CNR. The TOF CBCT method used in this work was verified with the same setup as in [14]. Then the TOF-CT method was applied on a small cylindrical water phantom to see the effects of object size in its performance. To explore the TOF CBCT application in breast imaging, the GATE simulation was repeated with a numerical breast phantom as the imaging object.

This paper is organized as follows. In section 2, we present the methods and the setup of the TOF-CT simulation using FEL and polychromatic x-ray sources. In section 3, preliminary results that confirm our method are presented, and study results obtained from the simulations follow. The paper concludes with discussions of the results.

## 2. Methods
### 2.1 GATE programming

The GATE software is a GEANT4 wrapper which utilizes the macro language to ease the learning curve of GEANT4 and allow GEANT4 to be more accessible to researchers [22]. The GATE simulations in this work were parallelized and executed with a custom bash script on a 20 CPU workstation. The simulation wait time varied for each imaging setup ranging from one to six weeks due to the different number of x-rays, object size, and detector size. All imaging setups acquired 360 projections. The physics lists enabled in the simulations consisted of the photoelectric effect, Compton scattering, and Rayleigh scattering which are the primary physics processes accounted for in medical imaging. The GATE software stores all output as ROOT files [23]. The necessary data from the ROOT output file was extracted using custom C++ code and processed in MATLAB. The extracted data consisted of the detector pixel number and the photon flight time. The detector temporal resolutions were applied after the simulations in MATLAB.

### 2.2 TOF method

The TOF method has been described previously by Rossignol et al and a similar approach will be reviewed here [14]. The TOF method was applied to the data in MATLAB after the GATE simulations. For scatter rejection based on the TOF of the x-rays, each detector pixel only accepts the x-ray if the flight time is within the data acquisition time window of that pixel. This allows for the rejection of most scattered x-ray photons. With the available flight time information, x-ray counts are accepted only if the following condition holds true:

$$t_{i,min} - \epsilon < t_{p,flight} < t_{i,max} + \omega \qquad (1)$$

$t_{i,min}$, $t_{i,max}$ are the calculated minimum and maximum acquisition times for the $i$th pixel based on the size of the pixel area and its corresponding distance to the source. $t_{p,flight}$ is the flight time of the $p$th x-ray. The $\epsilon$ parameter was set so that all first detected x-ray photons of each pixel are accepted. The $\omega$ parameter was set so that majority of x-rays are captured and accepted while neglecting x-rays that arrive to the detector much later. $\epsilon$ and $\omega$ parameters allow for a reasonable tuning of the acceptance window if the output rate of the source poses an issue for a faster imaging acquisition. The values of the $\epsilon$ and $\omega$ parameters are verified on five different detector areas to confirm that all severely scattered counts are ignored. The parameters are verified with pixels located at the center and along the central column and central row directions of the detector. The pixels within the magnified image of the object on the detector should be used. An example of the TOF scatter rejection method is shown in Fig. 1. The lower threshold, $\epsilon$, and upper threshold, $\omega$, correspond to the green line and red lines in the plot. The x-ray counts which are severely scattered are ignored for image reconstruction. For the TOF method to be implemented successfully, every pixel in the detector panel must receive sufficient photon counts. At least 10 registered x-rays per pixel were determined to be necessary.

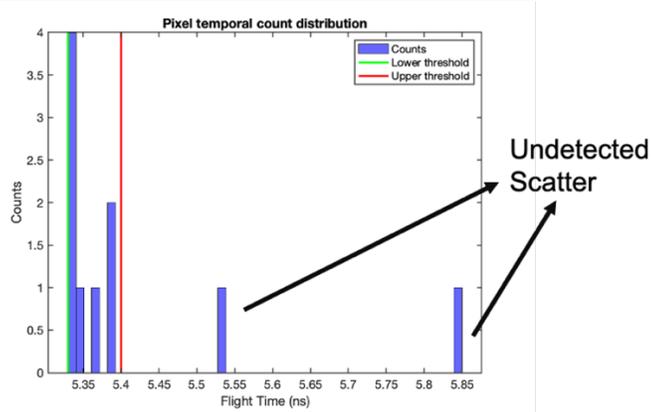

Figure 1: A typical example of the TOF scatter rejection method. $\epsilon$ and $\omega$ parameters (green and red lines respectively) were chosen so that counts that are severely scatted are not detected for image reconstruction.

*2.3 Setup geometry and data acquisition*

To verify the TOF method used in this work, a simulation study similar to Ref. 14 was performed. A large (24.4 cm diameter) cylindrical water phantom was simulated with two 3 cm diameter spine bone targets. The spine bone targets were inserted 80 mm apart. The phantom was placed at 1000 mm from the source. A flat panel detector was placed 1600 mm from the source. The detector had 512×512×1 silicon pixel elements with 1 mm$^3$ size. The detector temporal resolution was set to 10, 100, 500, and 1000 ps. A true 100 keV monoenergetic source was used to image the object. The source emitted photons with a 7° half cone angle. Due to the large size of the object and detector, $10^9$ x-ray photons were used per projection. Figure 2 shows the schematic of the imaging setup and a snapshot of the GATE simulation. For the GATE snapshot, a small sample of x-ray photons were initialized for viewing purposes.

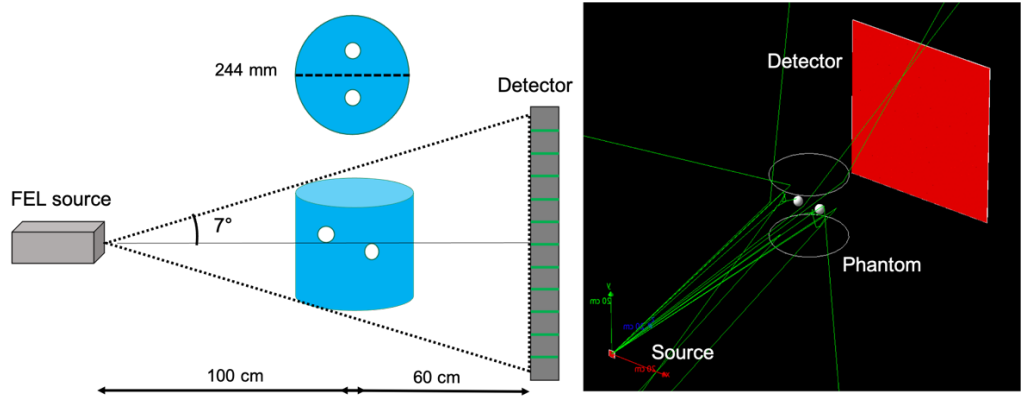

Figure 2: (left) Schematic of the imaging setup to validate the TOF method. An axial view of the object is included. The schematic is not drawn to scale. (right) Snapshot of the GATE simulation. The trajectories of the emitted x-rays are seen as green lines.

To image an object with an FEL source, a 20 keV quasi-monoenergetic FEL source was modeled using the linear interpolation user spectrum tool from GATE. The energy of the emitted photon is determined according to a probability distribution created by piecewise-linear interpolation between the energies provided. The modeled FEL source spectrum is plotted in Fig. 3.

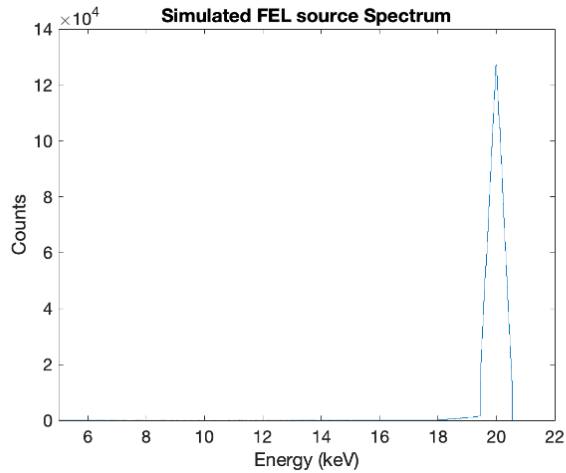

Figure 3: Spectrum of a modeled 20 keV quasi-monoenergetic FEL source

To explore the TOF methods on smaller objects with the modeled FEL source, an 80 mm diameter cylindrical water phantom with 5 mm diameter spine bone targets was used. The two targets were inserted 20 mm apart in the water phantom. Figure 4 shows the schematic of the TOF CBCT system used for this study. The phantom was positioned at 900 mm from the modeled FEL source. A 128×128×1 flat panel silicon detector was positioned at 1000 mm from the source. The pixel element size was set to 1 mm$^3$. The detector temporal resolution was set to 1, 2, 5, 10, 20, 50, 100, 200, 500 and 1000 picoseconds (ps) after the simulation. The photons were emitted from the source with a half angle of 4° in order to cover the imaging object. The detector was large enough that the base of the photon cone fits perfectly within the boundaries of the detector panel. $4 \times 10^7$ photons per projection were determined to be sufficient for this imaging setup while minimizing the simulation wait time. The $\epsilon$ and $\omega$ parameters for the TOF method for the small cylinder phantom were determined to be 3 and 8 ps, respectively.

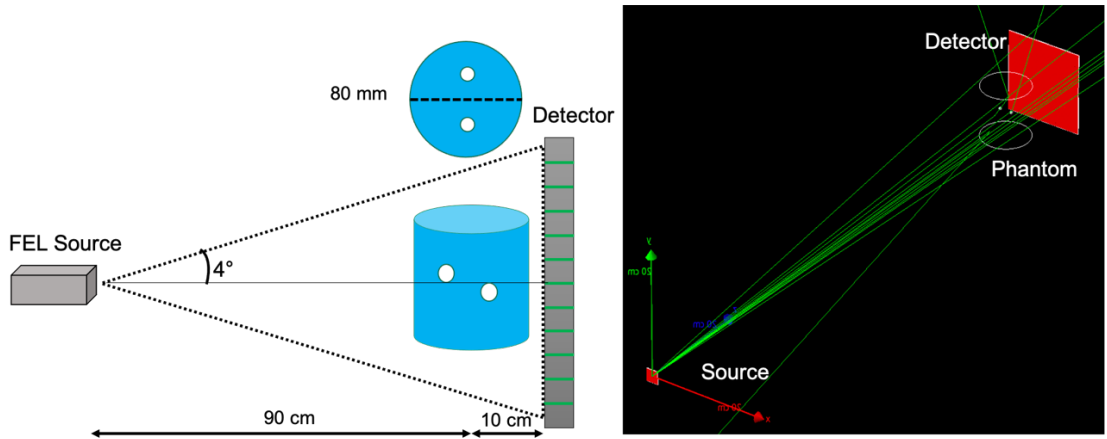

Figure 4: (left) Schematic of the TOF CBCT setup with a small cylinder phantom. An axial view of the imaging object is included. The schematic is not drawn to scale. (right) Snapshot of the GATE simulation. The trajectories of the emitted x-rays are seen as green lines.

For the TOF CBCT application in breast imaging, the same imaging setup as Fig, 4 was used. A breast phantom with a base diameter of 80 mm replaced the cylindrical water phantom. The breast phantom was generated using the VICTRE (Virtual Imaging Clinical Trials for Regulatory Evaluation) software [24] and imported into GATE using the nested parameterized method. The breast phantom was composed of adipose, blood, glandular, and muscle tissues. The breast phantom was discretized into 625×648×550 voxels. The voxel size was set to 0.128 mm$^3$. A 5248 $\mu$m diameter calcification composed of 8%wt calcium oxalate was integrated into the breast phantom as the target. Calcium oxalate has been found to exist at higher concentrations in malignant breast tissues and is used routinely to simulate calcifications in computer models [25-27]. Slices of the numerical breast phantom can be seen in Fig. 5. The dark regions of the breast are adipose tissues while the lighter regions are glandular tissues. The solid band along the left border of the image slices is muscular tissues resembling the start of the chest muscle. The calcification target is labeled by a yellow circle in breast slice 275 as show in Fig. 5b. The target was placed in the glandular tissues of the breast where development of malignancies is more prevalent. $3 \times 10^7$ photons per projection were determined to be sufficient. The remaining imaging parameters were the same as in the imaging of the small cylinder phantom. The same detector temporal resolutions were also applied. The $\epsilon$ and $\omega$ parameters for the TOF method were determined to be 3 and 3 ps respectively.

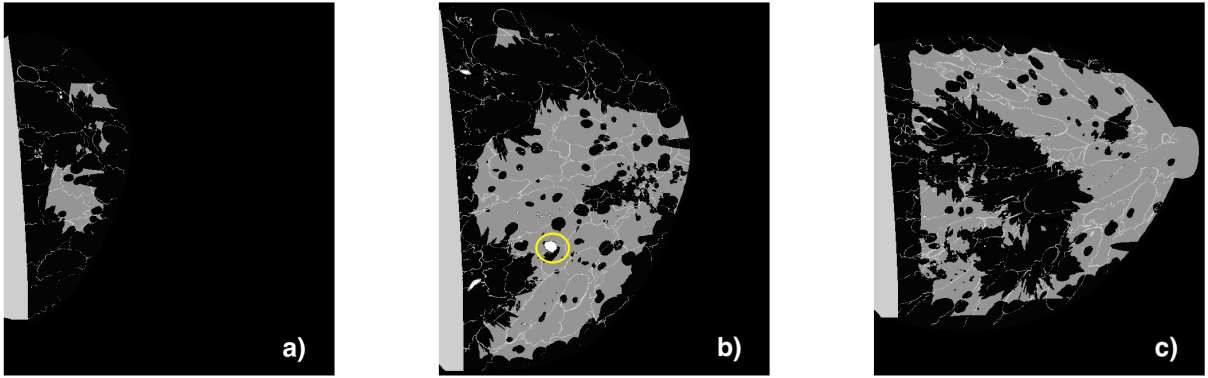

Figure 5: Representative slices of the breast phantom: a) slice 125, b) slice 275 which shows the inserted calcification with a yellow circle, and c) slice 450.

*2.4 Pre-processing and reconstruction algorithm*

The extracted data from the ROOT output was organized in a dynamic 3D array where the length of each element corresponded to the number of registered counts of each detector pixel with their respective TOF stamp. By using the TOF condition of Eq. (1), counts within each element vector were either accepted or rejected leading to different element vector lengths for every detector temporal resolution. The length of each array element for every detector temporal resolution led to a projection data set. Attenuation data sets were then calculated and fed to a Feldkamp (FDK) reconstruction algorithm.

The FDK reconstruction was performed using the Michigan Image Reconstruction Toolbox (MIRT) from Dr. Jeffrey Fessler from the University of Michigan in MATLAB [28]. For the preliminary study using the large cylindrical water phantom, the reconstruction voxel size was set to be 1 mm³. The reconstructed volume had dimensions of 512×512×512. For the small cylinder and breast phantoms, the reconstruction voxel size was set to be 0.75 mm³. The reconstructed volumes had dimensions of 128×128×128.

*2.5 Evaluation criteria*

Two criteria were used to evaluate the quality of the reconstructed images and the scatter reduction effectiveness of the TOF method.

CNR measurements were done using Eq. (2) to evaluate the image quality improvement based on the detector temporal resolution. The CNR was evaluated by using a square target region of interest (ROI) consisting of 9 image pixels.

$$CNR = \frac{\bar{x}_t - \bar{x}_{bkg}}{\sigma_{bkg}} \qquad (2)$$

$\bar{x}_t$ is the sample average pixel value from target ROI $t$, $\bar{x}_{bkg}$ and $\sigma_{bkg}$ are the background sample mean and standard deviation respectively [14]. The background image samples were an adjacent set of pixels from the calcification in the breast phantom and bone targets in the small cylinder phantom.

Scatter-to-primary ratio (SPR) measurements were performed using Eq. (3) to evaluate the scatter count reduction based on the detector temporal resolution [14].

$$SPR = \frac{N_{scatter}}{N_{primary}} \qquad (3)$$

$N_{scatter}$ are the number of scatter x-rays that violate the TOF condition in Eq. (1). $N_{primary}$ are the number of primary x-rays that meet the TOF condition. A 60×60 pixel region from the detector center was used for the SPR calculation. The registered photons in this detector region propagated through the object therefore obtaining primary and scattered photon counts for the SPR calculation.

## 3. Results
### 3.1 Preliminary results of the TOF method

The reconstructed TOF CBCT images of the large cylinder phantom using 10 ps and 1000 ps detector temporal resolutions are plotted in Fig. 6. The subfigures compare the reconstructed images with and without the TOF method. Fig. 6 also includes line profile plots showing the image intensity along the red and blue lines of the reconstructed images.

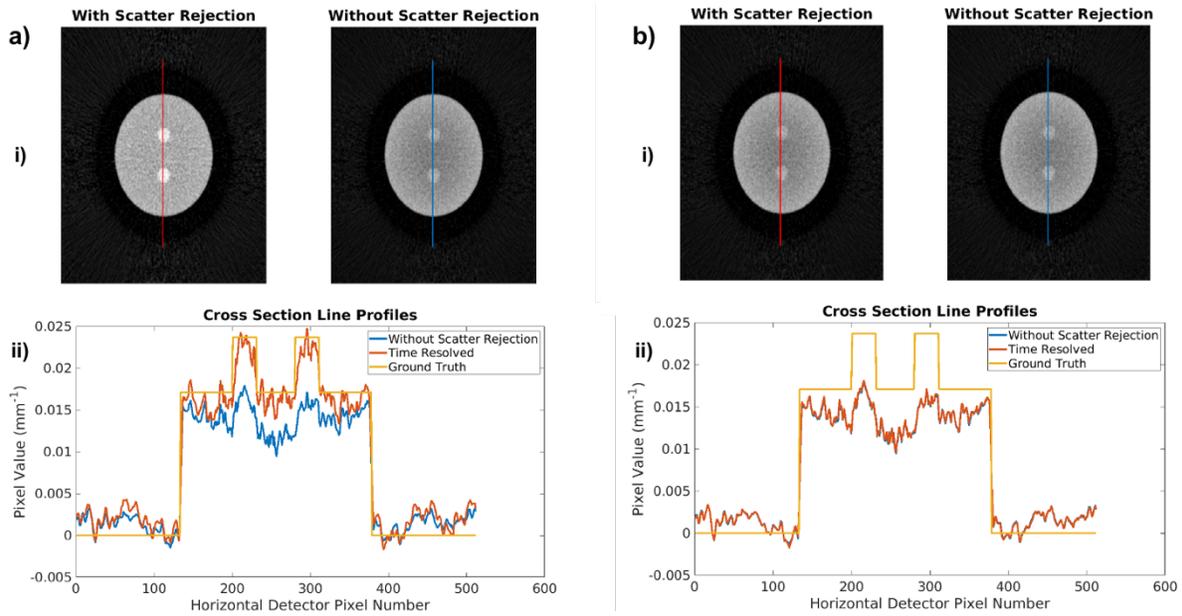

Figure 6: Preliminary results verifying the TOF method. ai) Reconstructed images with 10 ps detector temporal resolution (left) and without the TOF method (right). aii) Line profiles of the pixel intensity from images with 10 ps detector temporal resolution and without the TOF method. bi) Reconstructed images with 1000 ps detector temporal resolution (left) and without the TOF method (right). bii) Line profiles of the pixel intensity from images with 1000 ps detector temporal resolution and without the TOF method.

The subfigures on the left of Figs. 6ai) and 6bi) show reconstructed axial slices image of the phantom with the TOF method using 10 ps and 1000 ps temporal resolution respectively. The subfigures on the right were reconstructed without the TOF method. At 10 ps temporal resolution, a clear removal of the cupping effect was seen. From the line profile plot in Fig. 6aii), a general agreement between the ground truth and the TOF method was met. At 1000 ps resolution, the cupping artifacts remain apparent from the x-ray scattering events in the phantom. From the line profile plots in Figure 6bii), no significant improvement to the ground truth was achieved with 1000 ps resolution.

The scatter-to-primary ratio (SPR) plot for the large cylindrical water phantom is seen in Fig, 7. As the detector temporal resolution was increased, the SPR with TOF method approached the SPR without TOF method which was 1.23.

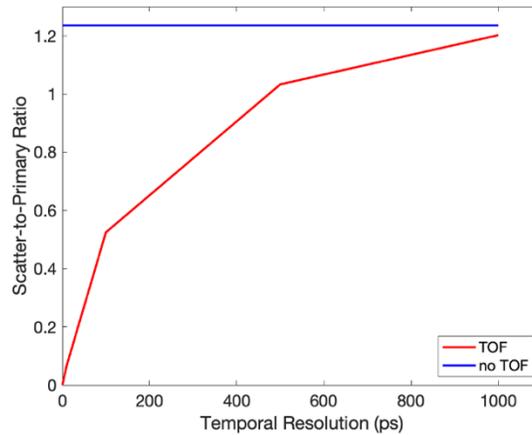

Figure 7: SPR plot of the preliminary TOF-CT study.

The CNR was improved by 1.4 at 100 ps detector temporal resolution and improved by 2 at 10 ps temporal resolution. The results obtained in this preliminary study were consistent with the results reported in Ref. 14.

*3.2 Results of the small cylinder phantom*

Fig. 8a shows the line profile of the small cylinder phantom for various detector temporal resolutions. Slight cupping artifacts were observed, but not as severe as in the imaging of the large cylinder phantom from the preliminary study. Target signal intensity improvement was seen when a temporal resolution of 20 ps was used. No further improvement in the target pixel intensity was seen with larger temporal resolutions than 10 ps. The reconstructed TOF CBCT images with 10 ps TOF scatter rejection and without scatter rejection are seen in Figure 8b. A clear increase in the contrast between background and targets is observed with 10 ps temporal resolution. The line profiles from Fig. 8a were generated from the intensities of the pixels along the yellow line of the reconstructed images for each detector resolution.

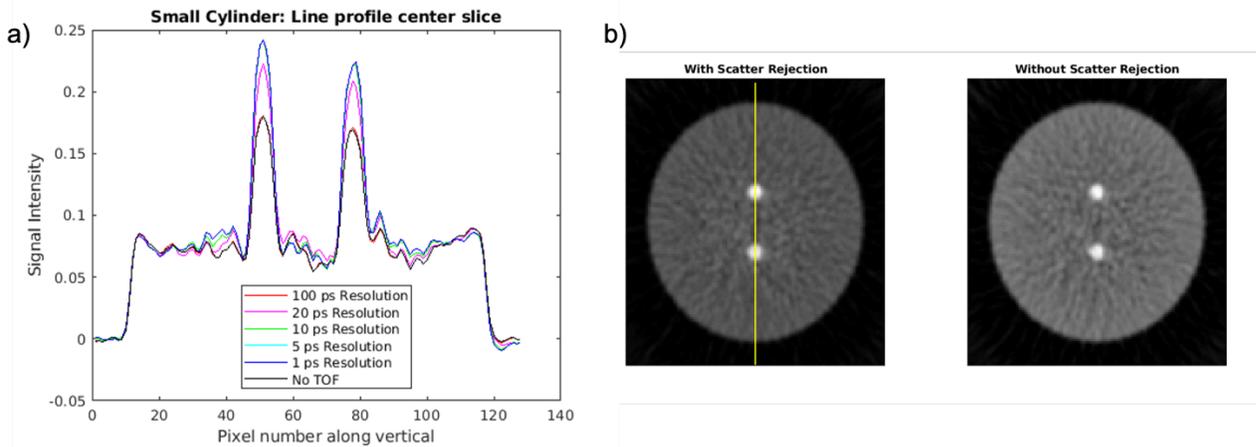

Figure 8: a) Line profiles of the small cylindrical phantom with various detector temporal resolutions. b) Cylinder slices of the reconstructed TOF CBCT images with 10 ps TOF scatter rejection and without TOF scatter rejection.

Fig. 9 shows the SPR plot for the small cylindrical phantom. The SPR without TOF scatter rejection was calculated to be 0.3401. The SPR values with TOF rejection reach the SPR without TOF rejection at 100 ps. This result also supports the results shown in the line profile plot from Fig. 8a, in which the signal intensity of the bone targets did not notably increase until detector temporal resolutions below 100 ps were used.

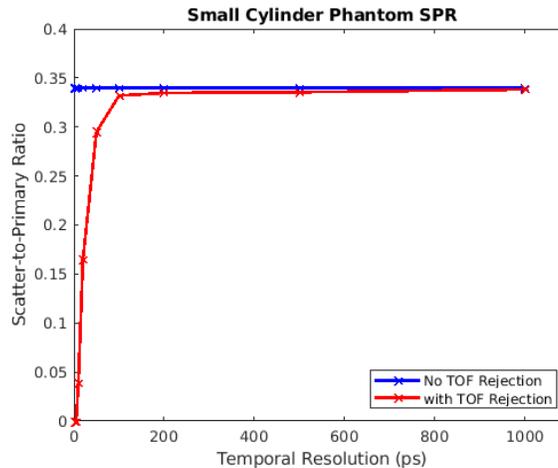

Figure 9: The SPR plot for the small cylindrical phantom case.

The CNR and SPR metrics for the small cylindrical phantom are shown in Table 1. The max CNR value with the TOF method was 25.071 with detector resolutions of 1, 2, and 5 ps. The lowest CNR was calculated to be 14.808 at 500 ps. The CNR is improved by a factor of 1.65 when the TOF method with a detector resolution of 1 to 5 ps was used. A 10 ps temporal resolution led to a CNR increase by a factor of 1.57. The lowest nonzero SPR value was calculated to be 0.0394 at 10 ps temporal resolution while the largest SPR value with TOF method was 0.3382 at 1000 ps. At 10 ps temporal resolution, the SPR was reduced by a factor of 8.63.

Table 1: CNR and SPR metrics for the varied detector temporal resolutions for the small cylinder phantom.

| Detector Temporal Resolution (ps) | CNR | SPR |
|---|---|---|
| 1 | 25.071 | 0.00 |
| 2 | 25.071 | 0.00 |
| 5 | 25.071 | 0.00 |
| 10 | 23.817 | 0.0394 |
| 20 | 17.585 | 0.1652 |
| 50 | 15.195 | 0.2949 |
| 100 | 15.470 | 0.3319 |
| 200 | 14.989 | 0.3347 |
| 500 | 14.808 | 0.3355 |
| 1000 | 15.178 | 0.3382 |
| No TOF rejection | 15.178 | 0.3401 |

*3.3 Results of breast phantom*

Slices of the reconstructed TOF CBCT images of the breast phantom are plotted in Fig. 10. The reconstructed calcification target can be seen in slice 65 as shown in Fig. 10b, in which the target was labeled by the yellow circle. An accurate reconstruction was obtained based on the preservation of the breast structures seen in Fig. 5.

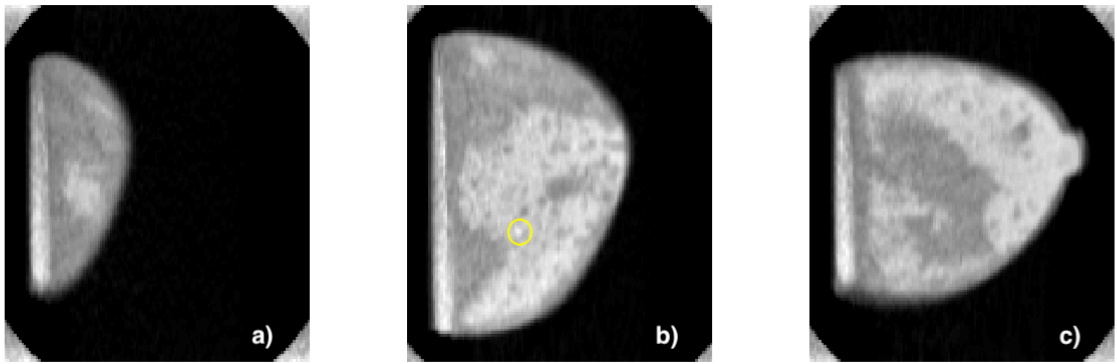

Figure 10: Slices of the reconstructed TOF CBCT images of the breast phantom: a) slice 35, b) slice 65 which shows the calcification targets labeled by the yellow circle, and c) slice 90.

Fig. 11a shows the line profile of the reconstructed images of the breast phantom for various detector temporal resolutions. The improvement of the reconstructed calcification target intensity was seen when a temporal resolution of 10 ps was used. However, the intensity improvement persisted for detector temporal resolutions of 5 ps and 1 ps. From the line profiles, the intensities from the surrounding glandular tissues were also enhanced by the TOF method, but not to the same extent as the calcification intensity. The adipose tissues remained relatively constant. Cupping artifacts were not apparent in the line profiles. The reconstructed images with 5 ps TOF scatter rejection and without TOF scatter rejection are plotted in Fig. 11b.

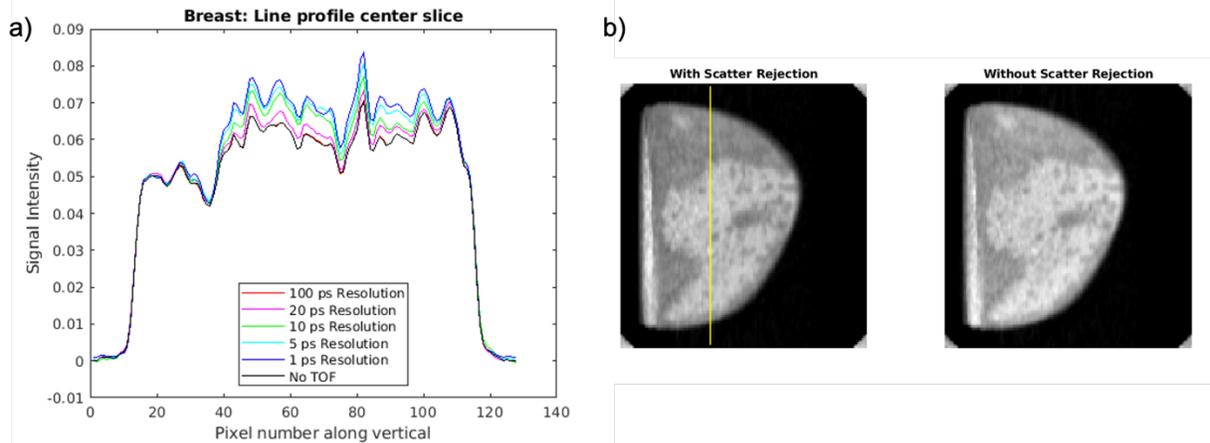

Figure 11: a) Line profiles of the reconstructed TOF CBCT images for the breast phantom with various detector temporal resolutions. b) Reconstructed breast images with 5 ps TOF scatter rejection and without TOF scatter rejection.

Fig. 12 shows the SPR plot for the breast phantom case. The plot was zoomed in along the x-axis to better show the curve changes with lower detector temporal resolutions. The SPR without TOF scatter rejection was 0.0998. The SPR curve with TOF rejection reached the SPR curve without TOF rejection at 100 ps which was similar to the small cylinder phantom SPR plot.

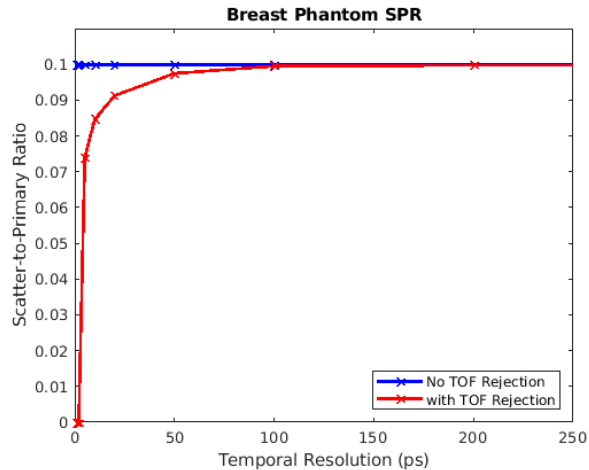

Figure 12: SPR plot of the breast phantom case.

The CNR and SPR metrics for the breast phantom case are shown in Table 2. The max CNR value with the TOF method was calculated to be 3.968 with detector resolutions 1 and 2 ps. The lowest CNR was calculated to be 3.522 at 100 ps. The CNR was improved by a factor of 1.13 when the TOF method with a detector resolution of 1 and 2 ps were used. A 5 ps temporal resolution led to a CNR increase by a factor of 1.10. The lowest nonzero SPR value was calculated to be 0.0739 at 5 ps temporal resolution while the largest SPR value with TOF method was 0.0998. At 5 ps temporal resolution, the SPR is reduced by a factor of 1.35.

Table 2: CNR and SPR metrics for the varied detector temporal resolutions for the breast phantom case.

| Detector Temporal Resolution (ps) | CNR | SPR |
|---|---|---|
| 1 | 3.968 | 0.0000 |
| 2 | 3.968 | 0.0000 |
| 5 | 3.944 | 0.0739 |
| 10 | 3.871 | 0.0847 |
| 20 | 3.714 | 0.0912 |
| 50 | 3.587 | 0.0975 |
| 100 | 3.527 | 0.0995 |
| 200 | 3.536 | 0.0997 |
| 500 | 3.521 | 0.0998 |
| 1000 | 3.522 | 0.0998 |
| No TOF rejection | 3.522 | 0.0998 |

## 4. Discussions and Conclusions

TOF-CT imaging with a modeled 20 keV FEL x-ray source and detector temporal resolutions of 1, 2, 5, 10, 20, 50, 100, 200, 500, and 1000 ps was performed in this work. The TOF method used in this work was verified using a cylindrical water phantom with bone targets as in Ref. 14. For the small cylinder phantom, $4 \times 10^7$ photons per projection were used while $3 \times 10^7$ x-rays per projection were used for the breast phantom to minimize the wait time while providing sufficient counts to every detector pixel element. The difference in the number of initialized photons arise from the different geometries between the two objects. The breast phantom resembles a cone-shaped object more than a cylindrical object, so more photons are expected to reach the detector with the breast phantom. Therefore, to keep the simulation wait time relatively shorter, the number of photons were reduced for the breast phantom.

Due to the prolonged simulation wait time, the number of x-ray photons per projection, pixel number of the detector, and size of the breast phantom were limited. Better statistics would be acquired with a greater number of x-ray photons and greater number of detector pixel element which would improve the values seen in Table 1 and Table 2 such that nonzero SPR values would be observed for all detector temporal resolutions. A smoother trend in CNR values with respect to detector temporal resolutions would also be observed. The slight variation of CNR values is due to the image noise. Nonetheless the general trend in SPR and CNR metrics agree with theory. The TOF effects would be better demonstrated on a thicker breast phantom of 10-14 cm in diameter. At this size, greater scattering effects will take place due to the irradiation of a larger volume by x-rays such that greater temporal resolutions will suffice to improve the CNR and SPR significantly.

A 10 ps detector temporal resolution led to a 1.57 improvement in target CNR and 8.63 reduction in SPR for the small cylinder phantom. A detector temporal resolution below 5 ps was needed to see a notable enhancement in CNR and reduction in SPR for the calcification target for the breast phantom. Due to the differences in geometry and phantom composition, significant discrepancies in the SPR and CNR were shown. The x-rays are forced to travel through more matter in the cylindrical phantom, especially along the center of the object. Therefore, a greater SPR is apparent for the smaller cylinder phantom than the breast phantom. The SPR results between the small cylinder and breast phantoms may be more agreeable if a cone shaped phantom was imaged instead of a cylindrical phantom.

The differences in the CNR enhancement between the small cylinder and breast phantom arise from the differences in the target composition. The photoelectric effect is responsible for the differences in contrast between different materials. The bone target exhibits much higher attenuation due to a higher effective atomic number than the calcification composed of 8%wt calcium oxalate. With similar background compositions in the breast and small cylinder phantoms, the CNR was expected to be much greater for the small cylinder phantom.

Currently, the TOF method will not be suitable for implementation in cone beam breast imaging due to the limitations of modern detector technology. From the results for the breast phantom, a detector resolution below 10 ps was required to see notable enhancements in the image quality and scatter reduction. However, with rapid advancements in detector technology, the TOF method may become available in the future for dedicated breast cone beam CT systems which will replace conventional mammogram systems.

In this work, an ideal detector was used to remove detector noise factors. The goal of this study was to theoretically explore the TOF method in smaller objects therefore the detector noise was factored out by considering an ideal detector. All detector temporal resolution variations were applied after the GATE simulations in MATLAB. To translate the methods of this work to the experimental setting successfully, the detector noise will need to be minimized.

In summary, the TOF method was implemented on cone beam CT imaging of a small cylinder phantom and numerical breast phantom. A set of GATE simulations were performed to study the TOF method effectiveness of these small objects. Our simulation study indicates that a TOF resolution below 10 ps was required to see notable enhancements in the image quality and scatter reduction for breast TOF CBCT imaging. The results of this work will propel the future exploration of the TOF CBCT imaging and its applications.